\documentclass[12pt]{article}
\usepackage{graphicx}
\usepackage[]{amsmath}
\usepackage{lineno}
\usepackage{xspace}
\usepackage{color}
\usepackage{subcaption}

\newcommand\pubnumber{CIPANP2018-Rettie}
\newcommand\pubdate{\today}

\newcommand*{\pt}{\ensuremath{p_{\text{T}}}\xspace}
\newcommand*{\mT}{\ensuremath{m_{\text{T}}}\xspace}
\newcommand*{\met}{\ensuremath{E_{\text{T}}^{\text{miss}}}\xspace}

\newcommand*{\TeV}{\ensuremath{\text{Te\kern -0.1em V}}}
\newcommand*{\GeV}{\ensuremath{\text{Ge\kern -0.1em V}}}
\newcommand*{\MeV}{\ensuremath{\text{Me\kern -0.1em V}}}
\newcommand*{\keV}{\ensuremath{\text{ke\kern -0.1em V}}}
\newcommand*{\eV}{\ensuremath{\text{e\kern -0.1em V}}}

\newcommand*{\ifb}{\mbox{fb$^{-1}$}}

\newcommand*{\fig}[4][0.6]{ 
	\begin{figure}
		\centering
		\includegraphics[width=#1\textwidth]{#2}
		\caption{#3}
		\label{#4}
	\end{figure}
 }
\newcommand*{\doublefig}[6]{ 
	\begin{figure}
	\centering
		\begin{subfigure}{0.49\linewidth}
			\centering
			\includegraphics[width=\textwidth]{#1}
			\caption{\label{#2}}
		\end{subfigure}
		\begin{subfigure}{0.49\linewidth}
			\centering
			\includegraphics[width=\textwidth]{#3}
			\caption{\label{#4}}
		\end{subfigure}
		\caption{#5}
		\label{#6}
	\end{figure}
 }

\def\UBCTRIUMF{The University of British Columbia -- TRIUMF}
\def\support{\footnote{The author acknowledges support from the Vanier Canada Graduate Scholarship program, and the Natural Sciences and Engineering Research Council of Canada.\\\\Copyright 2018 CERN for the benefit of the ATLAS Collaboration. Reproduction of this article or parts of it is allowed as specified in the CC-BY-4.0 license.}}

\def\Title#1{\begin{center} {\Large #1 } \end{center}}
\def\Author#1{\begin{center}{ \sc #1} \end{center}}
\def\Address#1{\begin{center}{ \it #1} \end{center}}

\newcommand\pubblock{\rightline{\begin{tabular}{l} \pubnumber\\
         \pubdate  \end{tabular}}}
\newenvironment{Abstract}{\begin{quotation}  }{\end{quotation}}
\newenvironment{Presented}{\begin{quotation} \begin{center} 
             PRESENTED AT\end{center}\bigskip 
      \begin{center}\begin{large}}{\end{large}\end{center} \end{quotation}}

\def\beq{\begin{equation}}
\def\eeq#1{\label{#1}\end{equation}}
\def\eeqn{\end{equation}}

\def\beqa{\begin{eqnarray}}
\def\eeqa#1{\label{#1}\end{eqnarray}}
\def\eeqan{\end{eqnarray}}

\let\bar=\overbar

\def\Dslash{\not{\hbox{\kern-4pt $D$}}}
\def\dslash{\not{\hbox{\kern-2pt $\del$}}}

\def\msb{{\bar{\ssstyle M \kern -1pt S}}}

\begin{document}
\begin{titlepage}
\pubblock

\vfill
\Title{Searches for new phenomena in leptonic final states using the ATLAS detector}
\vfill
\Author{Sébastien Rettie\support, on behalf of the ATLAS Collaboration}
\Address{\UBCTRIUMF}
\vfill
\begin{Abstract}
Many theories beyond the Standard Model predict new phenomena which decay to well isolated, high-\pt{} leptons. Searches for new physics with these signatures are performed using the ATLAS experiment at the LHC. The results reported here use the pp collision data sample collected by the ATLAS detector at the LHC with a center-of-mass energy of 13~\TeV{}.
\end{Abstract}
\vfill
\begin{Presented}
Thirteenth Conference on the Intersections of Particle and Nuclear Physics\\
Palm Springs, CA, May 28 -- June 3, 2018
\end{Presented}
\vfill
\end{titlepage}
\def\thefootnote{\fnsymbol{footnote}}
\setcounter{footnote}{0}

\section{Introduction}
The Standard Model (SM) of particle physics is a very successful predictive theory which explains the fundamental interactions of elementary particles in the universe, except for gravity. However, the SM is known to be an effective theory that is valid only in a low energy regime, called the electroweak scale, and does not account for many observed experimental results. For example, it does not offer a satisfying explanation for neutrino masses or dark matter. Hence, it is clear that to fully understand and explain nature, a theoretical framework that goes beyond the Standard Model (BSM) is required.

The ATLAS experiment~\cite{Aad:2008zzm} at the Large Hadron Collider (LHC) has collected an unprecedented amount of data at $\sqrt{s}=13~\TeV{}$. This dataset can be used to search for new physics in various ways. Leptons are very clean objects within the ATLAS detector, and can be triggered on quite efficiently. Hence, searching for new physics in events with leptonic final states is extremely appealing. Using the latest ATLAS data, searches for new phenomena in leptonic final states are performed. In particular, section~\ref{Wprime_lnu} presents a search for a new heavy gauge boson resonance decaying into a light lepton (electron or muon) and a neutrino, section~\ref{Wprime_taunu} presents a search for high-mass resonances decaying into a tau and a neutrino, section~\ref{Zprime} presents a search for new high-mass phenomena with two light leptons in the final state, and section~\ref{LFVZ} presents a search for lepton-flavor-violating decays of the Z boson into a tau and a light lepton.


\section{$W' \rightarrow l\nu$ search}
\label{Wprime_lnu}

To isolate the potential signal coming from the decay of a new heavy gauge boson resonance decaying into a light lepton (electron or muon) and a neutrino~\cite{Wprime}, events with large missing transverse momentum (\met{}) and containing exactly one isolated lepton with large transverse momentum (\pt{}) are selected. The discriminating variable used in this $W' \rightarrow l\nu$ search is the transverse mass 

\begin{equation}
	\mT{} = \sqrt{2\pt{}\met{}(1-\cos\phi_{l\nu})},
	\label{eq:mT}
\end{equation}

where $\phi_{l\nu}$ is the azimuthal angle between the directions of the lepton \pt{} and the \met{} in the transverse plane. The \mT{} distributions for the electron and the muon channel are shown in Figure~\ref{fig:Wprime_mT}. The multi-jet background process is estimated using the data-driven matrix method, whereas all other background processes are estimated using Monte Carlo (MC). The dominant background process is the charged-current Drell-Yan (DY) W boson production (shown in white in Figure~\ref{fig:Wprime_mT}). The largest systematic uncertainties arise from the fake electron background estimation, the muon reconstruction, the lepton momentum scale and resolution, and the parton distribution function (PDF) variation.

In the absence of a new physics signal in the \mT{} spectrum, an upper limit is placed on the $W'$ cross-section ($\sigma$) times branching ratio ($BR$) into an electron or muon as a function of the $W'$ mass, as shown in Figure~\ref{fig:Wprime_limit_l}. From this, a lower limit on the mass of the Sequential Standard Model (SSM)~\cite{SSM} $W'$ boson, the benchmark new physics model used in this search, is placed at 5.6~\TeV{}. This mass limit corresponds to a 500~\GeV{} improvement with respect to the previous ATLAS result~\cite{Wprime_20152016}, which used only 36~\ifb{} of data compared to the 79.8~\ifb{} used for this analysis.

\doublefig
	{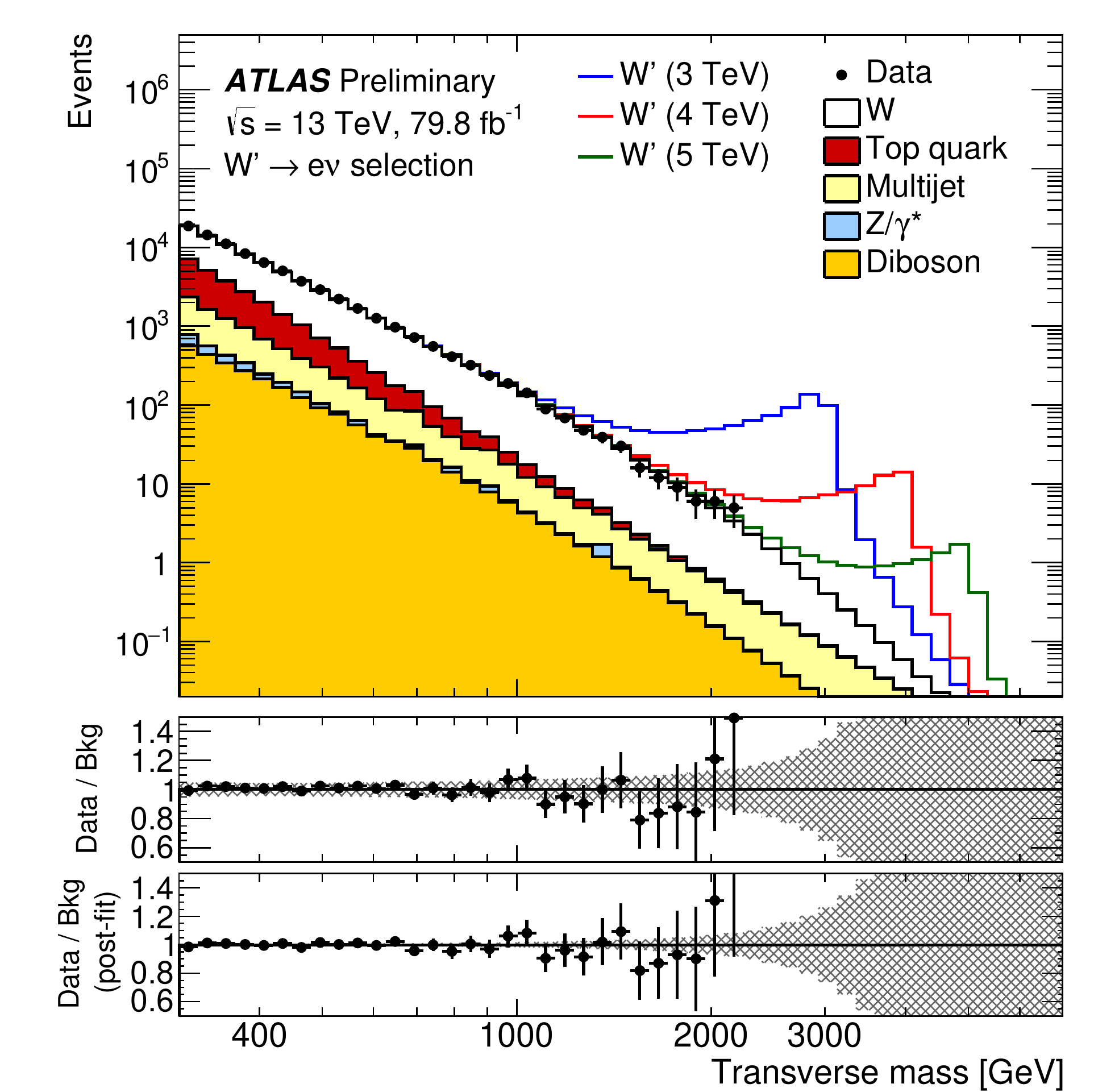}
	{fig:Wprime_mT_e}
	{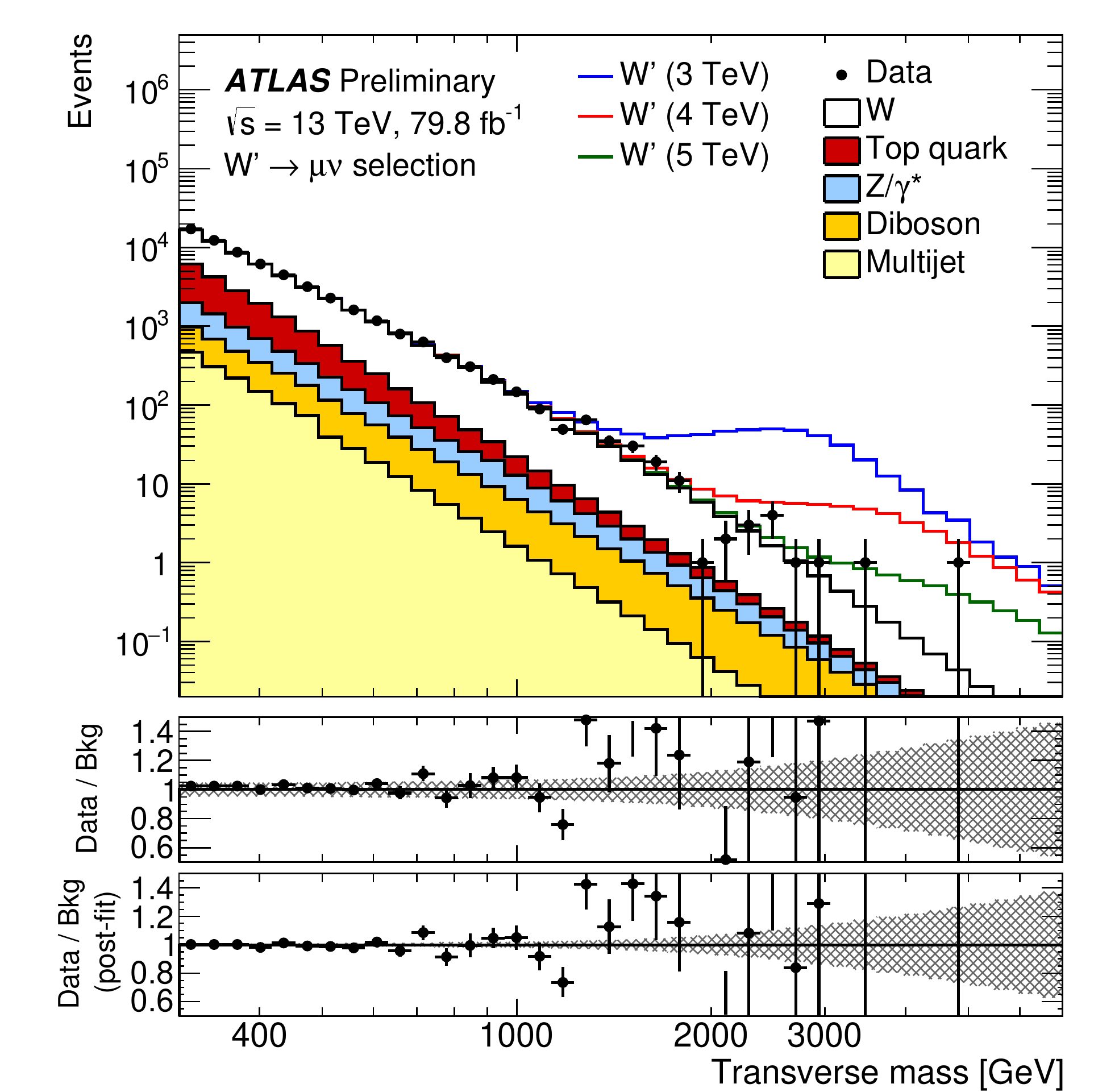}
	{fig:Wprime_mT_mu}
	{Transverse mass distribution after event selection for the (\subref{fig:Wprime_mT_e}) electron and (\subref{fig:Wprime_mT_mu}) muon channel of the $W' \rightarrow l\nu$ search~\cite{Wprime}.}
	{fig:Wprime_mT}

\fig[0.5]
	{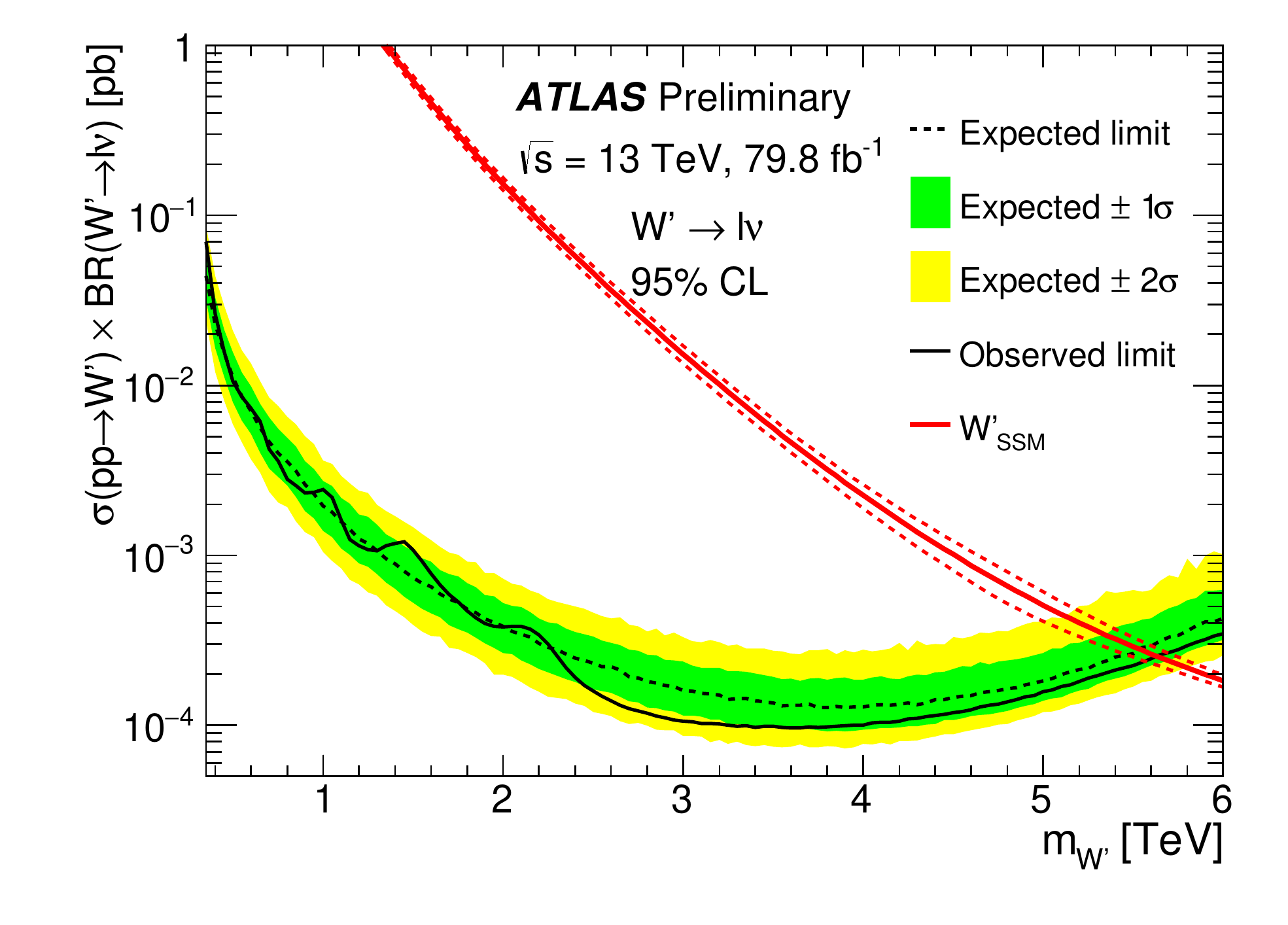}
	{95\% C.L. upper limit on $\sigma BR$ as a function of mass for the $W' \rightarrow l\nu$ search~\cite{Wprime}.}
	{fig:Wprime_limit_l}

\section{$W' \rightarrow \tau\nu$ search}
\label{Wprime_taunu}

$W'$ bosons can also decay into $\tau$ leptons and a neutrino. This decay mode is treated separately from the light lepton decays because of the different reconstruction techniques required for the $\tau$ lepton. The $\tau$ lepton decays into a W$^*$ boson and a neutrino, with the W$^*$ boson subsequently decaying into various final state particles. Only hadronically decaying $\tau$ leptons, which account for $\sim 65\%$ of the total $\tau$ lepton branching fraction, are considered in this analysis. Typically, hadronically decaying $\tau$ leptons decay into a neutrino, one or three charged pions, and up to two neutral pions. The search for high-mass resonances decaying into a tau lepton and a neutrino~\cite{Wprimetau} is particularly important for BSM models with enhanced third generation couplings, e.g. the non-universal G(221) model~\cite{g221}.

Events are required to have fired a \met{} trigger, and an offline \met{} cut of 150~\GeV{} is applied in order to minimize the trigger efficiency uncertainty. The visible $\tau$ lepton decay products candidate, $\tau_{\text{had-vis}}$, must have $\pt{} > 50~\GeV{}$, $|\eta| < 2.4$, one or three associated tracks, and net electric charge of $\pm1$. Finally, an electron and muon veto is imposed to reduce electrons misidentified as $\tau_{\text{had-vis}}$.

The discriminating variable used in this $W' \rightarrow \tau\nu$ search is the transverse mass of the $W'$ boson, as described in equation~\ref{eq:mT}, with the light lepton being replaced by the $\tau$ lepton. The \mT{} distribution of events passing the final selection is shown in Figure~\ref{fig:Wprimetau_mT}. The jet background is estimated using the data-driven fake factor method, and the non-jet background is estimated using MC simulation. By performing a counting experiment in the tail of the \mT{} distribution, limits on various BSM theory parameters can be set. Figure~\ref{fig:Wprimetau_limit} shows the upper limit on the cross-section times branching ratio of the $W'$ as a function of its mass. Figure~\ref{fig:Wprimetau_limit_cot_vs_mass} highlights the complementarity of this search to other BSM searches, extending the sensitivity over a large fraction of the $\cot\phi_{\text{NU}} - \text{m}_{W'}$ parameter space.

\fig[0.57]
	{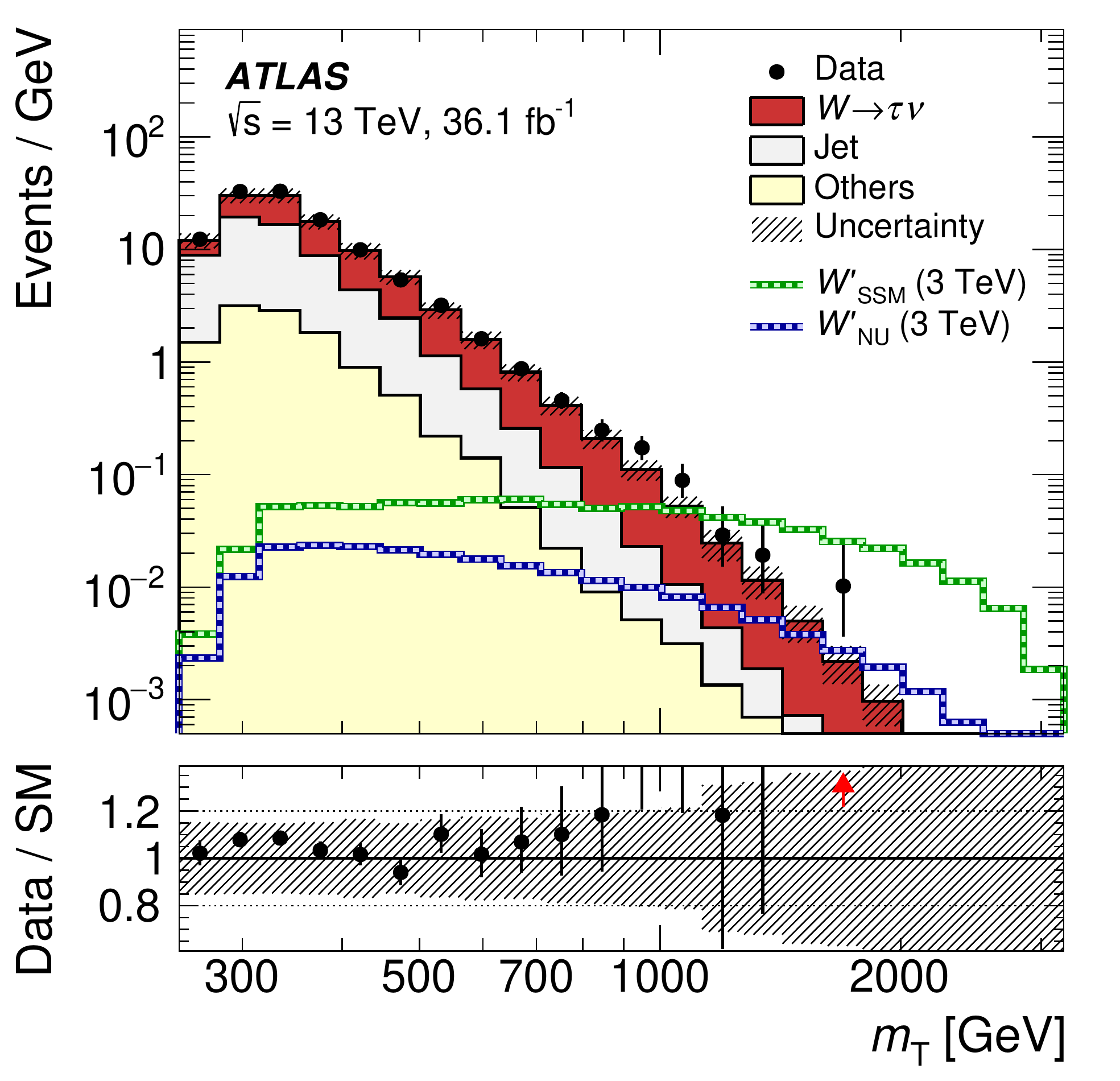}
	{Transverse mass distribution after event selection for the $W' \rightarrow \tau\nu$ search~\cite{Wprimetau}.}
	{fig:Wprimetau_mT}

\doublefig
	{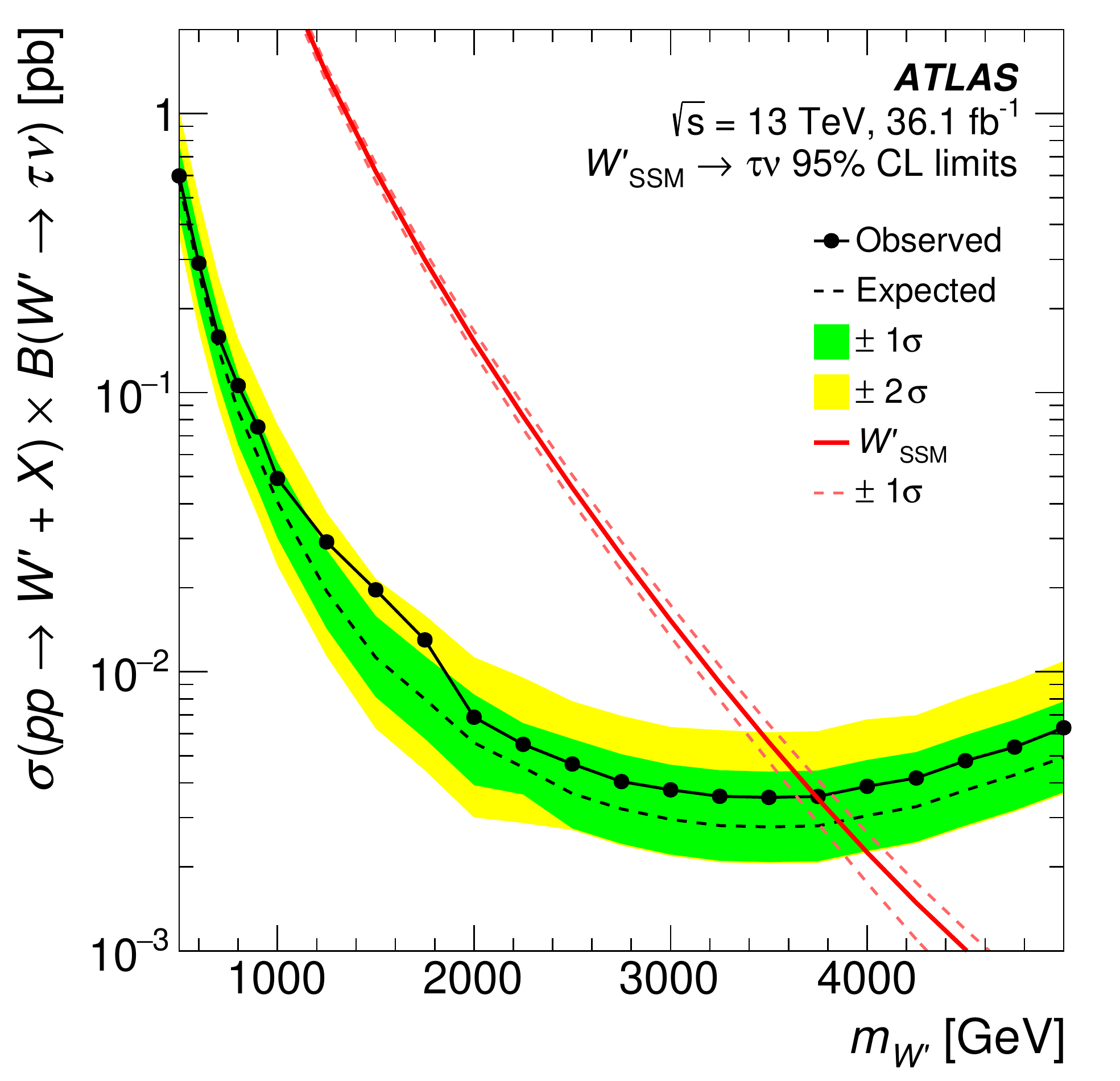}
	{fig:Wprimetau_limit}
	{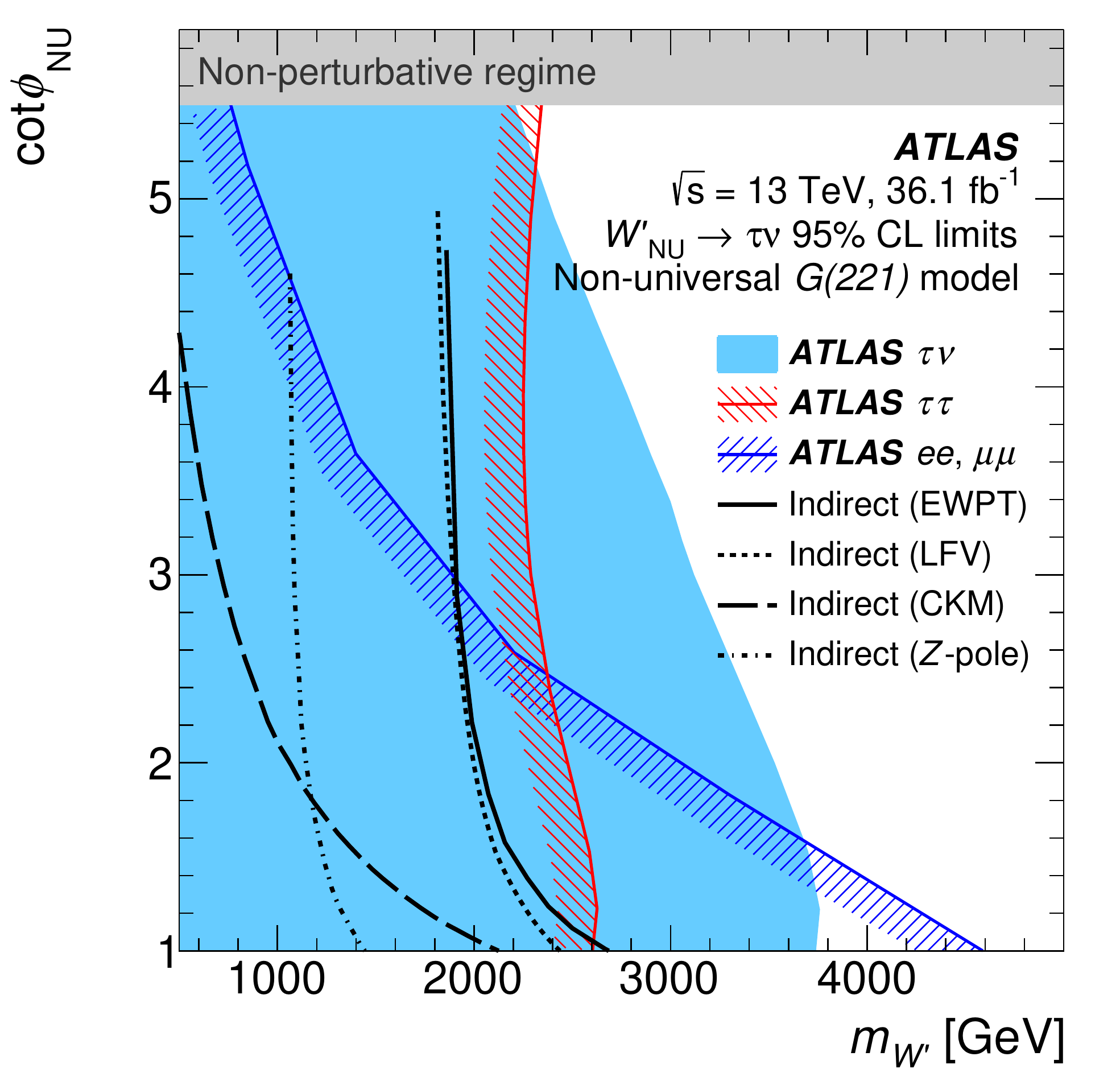}
	{fig:Wprimetau_limit_cot_vs_mass}
	{(\subref{fig:Wprimetau_limit}) 95\% C.L. upper limit on $\sigma B$ as a function of mass for the $W' \rightarrow \tau\nu$ search. (\subref{fig:Wprimetau_limit_cot_vs_mass}) 95\% C.L. exclusion regions in the $\cot\phi_{\text{NU}} - \text{m}_{W'}$ parameter space of the $W'_\text{NU}$ model~\cite{Wprimetau}.}
	{fig:Wprime_limits}

\newpage
\section{$Z' \rightarrow ll$ search}
\label{Zprime}

New high-mass phenomena such as $Z'$ bosons decaying into two light leptons in the final state are searched for by selecting events with one pair of same-flavor, isolated leptons with $\pt{} > 30~\GeV{}$~\cite{Zprime}. The dominant background is the neutral-current DY process which is estimated with MC, while the multi-jet and W+jets backgrounds in this search are estimated using the data-driven matrix method in the electron channel, and are found to be negligible in the muon channel.

The invariant mass of the dilepton system is used as the discriminating variable in this analysis, and its distribution is shown for the dielectron and dimuon channels in Figure~\ref{fig:Zprime_InvMass}. Upper limits on the $\sigma B$ of the $Z'$ boson as a function of its mass are shown in Figure~\ref{fig:Zprime_limit_ll}. Lower limits on the mass of the $Z'$ boson for various BSM models are also shown. In particular, a lower limit on the mass of the benchmark $Z'_{\text{SSM}}$ boson is set at 4.5~\TeV{}.

\doublefig
	{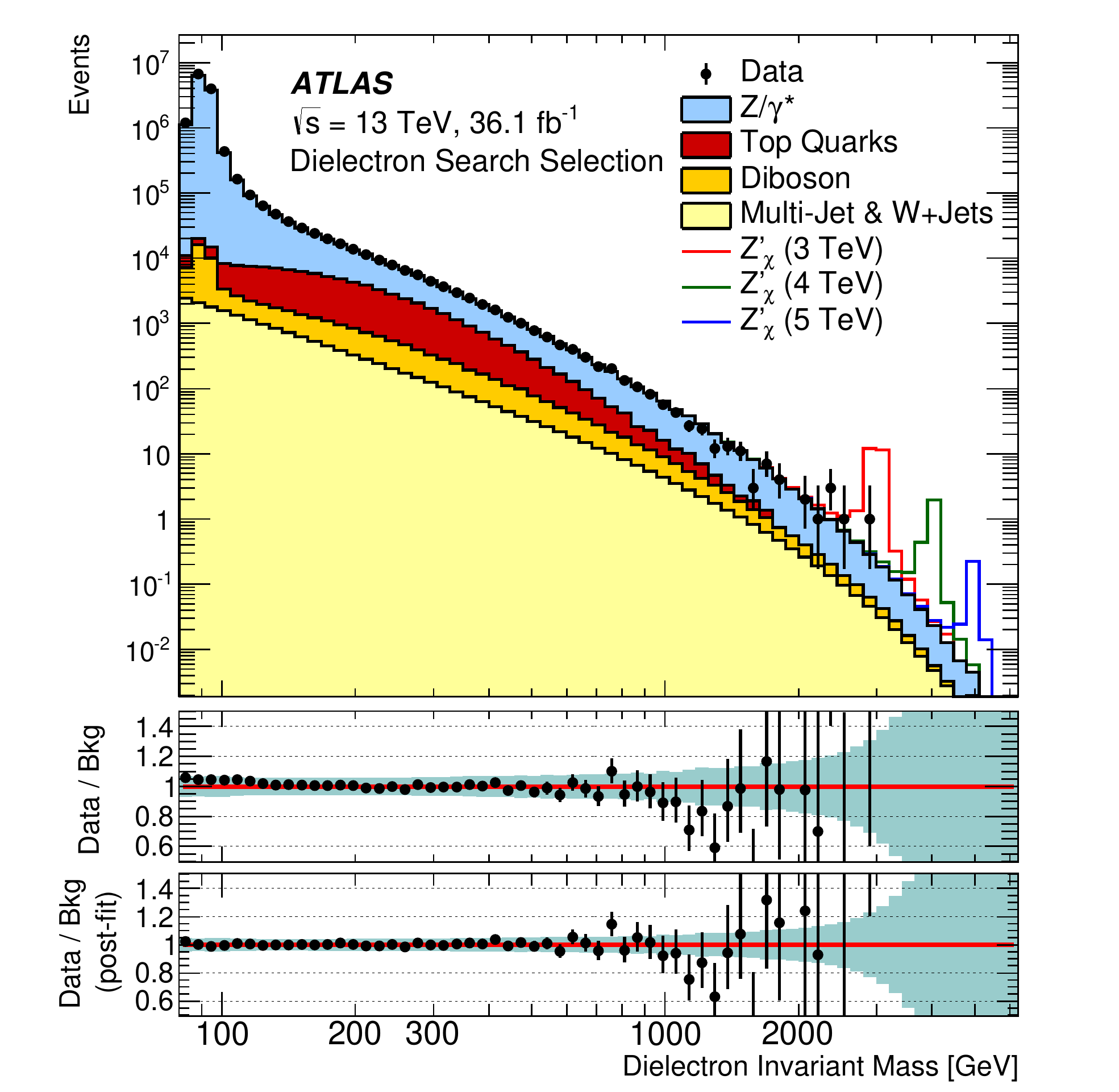}
	{fig:Zprime_m_ee}
	{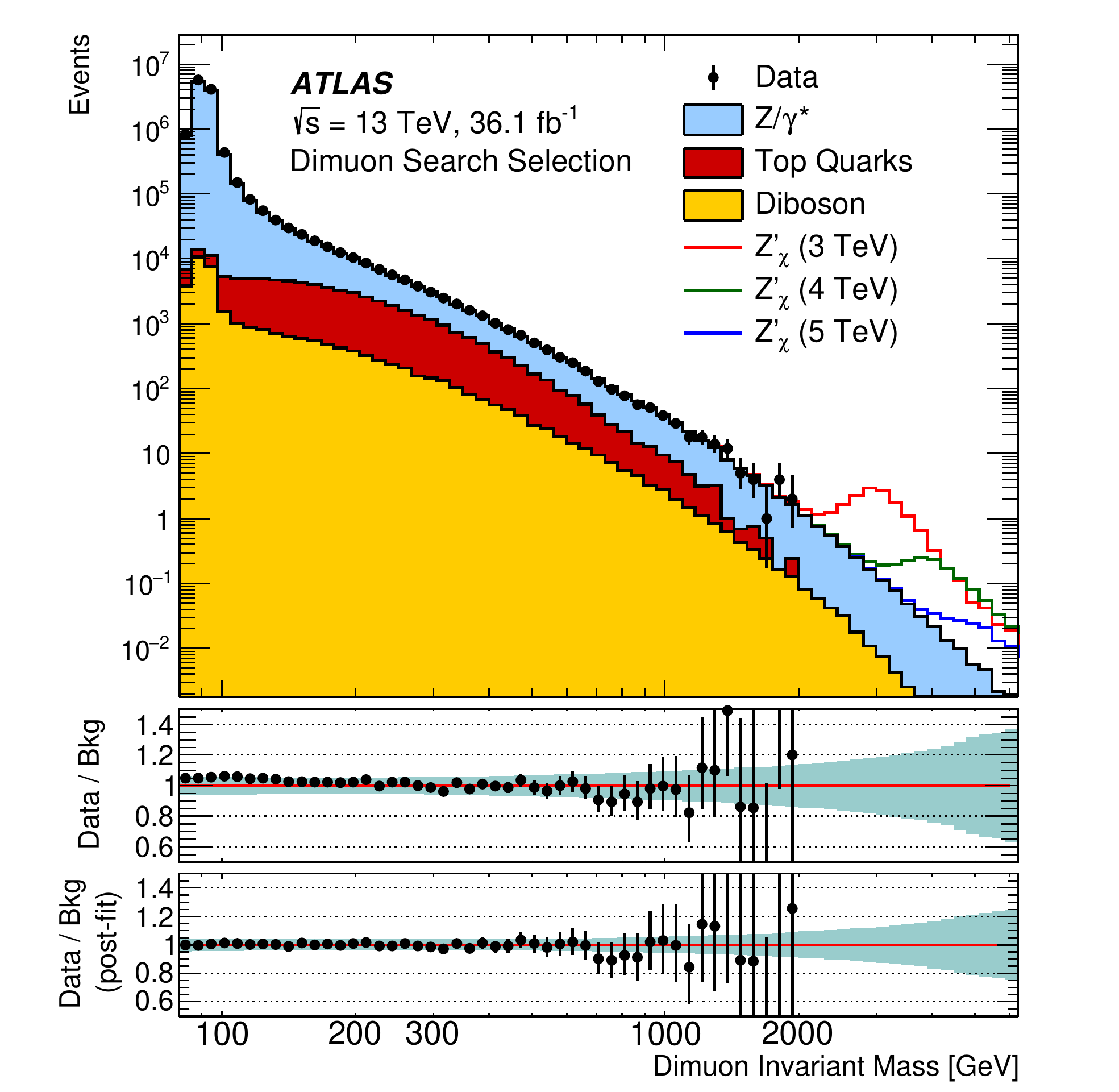}
	{fig:Zprime_m_mumu}
	{Invariant mass distributions for the (\subref{fig:Zprime_m_ee}) dielectron and (\subref{fig:Zprime_m_mumu}) dimuon channels of the $Z' \rightarrow ll$ search~\cite{Zprime}.}
	{fig:Zprime_InvMass}

\fig[0.57]
	{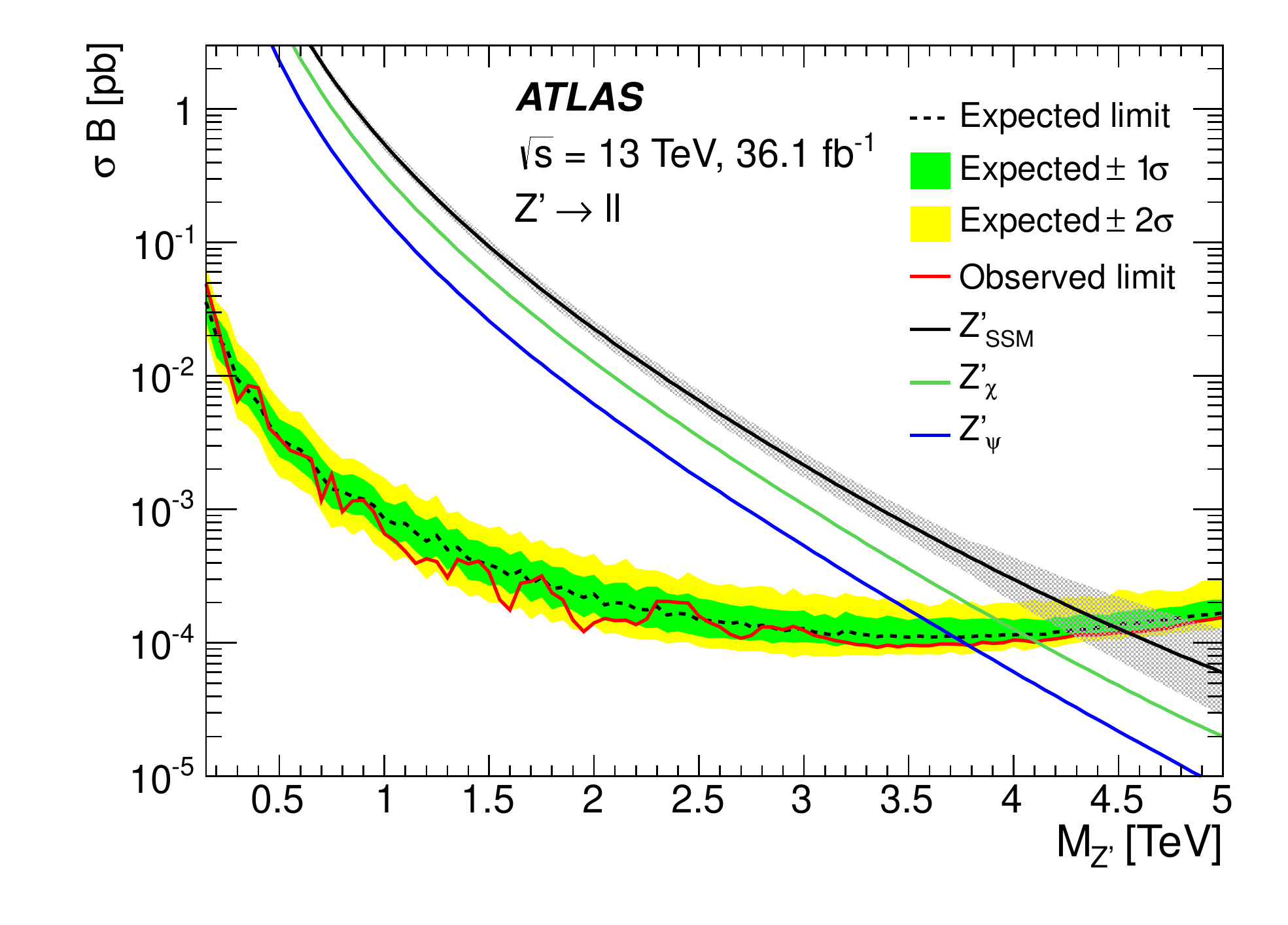}
	{95\% C.L. upper limit on $\sigma B$ as a function of mass for the $Z' \rightarrow ll$ search~\cite{Zprime}.}
	{fig:Zprime_limit_ll}

\section{Lepton-flavor-violating $Z$ boson search}
\label{LFVZ}

Lepton-flavor-violating (LFV) Z boson decays are predicted by models with heavy neutrinos~\cite{LFVZ_HeavyNeutrinos}, extended gauge models~\cite{LFVZ_ExtendedGauge}, etc. This search considers LFV decays of the Z boson into a light lepton (electron or muon) and a tau. Similarly to the search presented in section~\ref{Wprime_taunu}, this analysis only considers hadronically decaying $\tau$~\cite{LFVZ}. A cut on the invariant mass of the system composed of the light lepton and the $\tau_{\text{had-vis}}$ candidate or track is placed, rejecting events where the invariant mass is compatible with the Z boson mass in order to reject SM Z boson events (depicted in Figure~\ref{fig:LFVZ_cuts}). Signal events are expected to have the missing transverse momentum from the neutrino in a direction close to the $\tau_{\text{had-vis}}$ candidate, resulting in a small transverse mass between the $\tau_{\text{had-vis}}$ candidate and the \met{}. Figure~\ref{fig:LFVZ_mT_mutau} highlights this by comparing the \mT{} distributions for the signal and background processes of this analysis. Events where the $\tau_{\text{had-vis}}$ candidate originates from a quark- or gluon-initiated jet are estimated using the data-driven fake factor method, and all other processes are estimated from MC simulation. Finally, signal events passing the selection are classified using a neural network (NN) trained to discriminate $Z \rightarrow l\tau$ signal events from background events. The output distribution of the NN for the muon channel is shown in Figure~\ref{fig:LFVZ_NNout_mutau}. These output distributions are analyzed in a template fit to data, and upper limits on the LFV decay branching fraction are then set at the 95\% C.L. using the CL$_\text{S}$ method: $\mathcal{B}(Z \rightarrow \mu\tau) < 2.4 \times 10^{-5}$ and $\mathcal{B}(Z \rightarrow e\tau) < 5.8 \times 10^{-5}$.

\doublefig
	{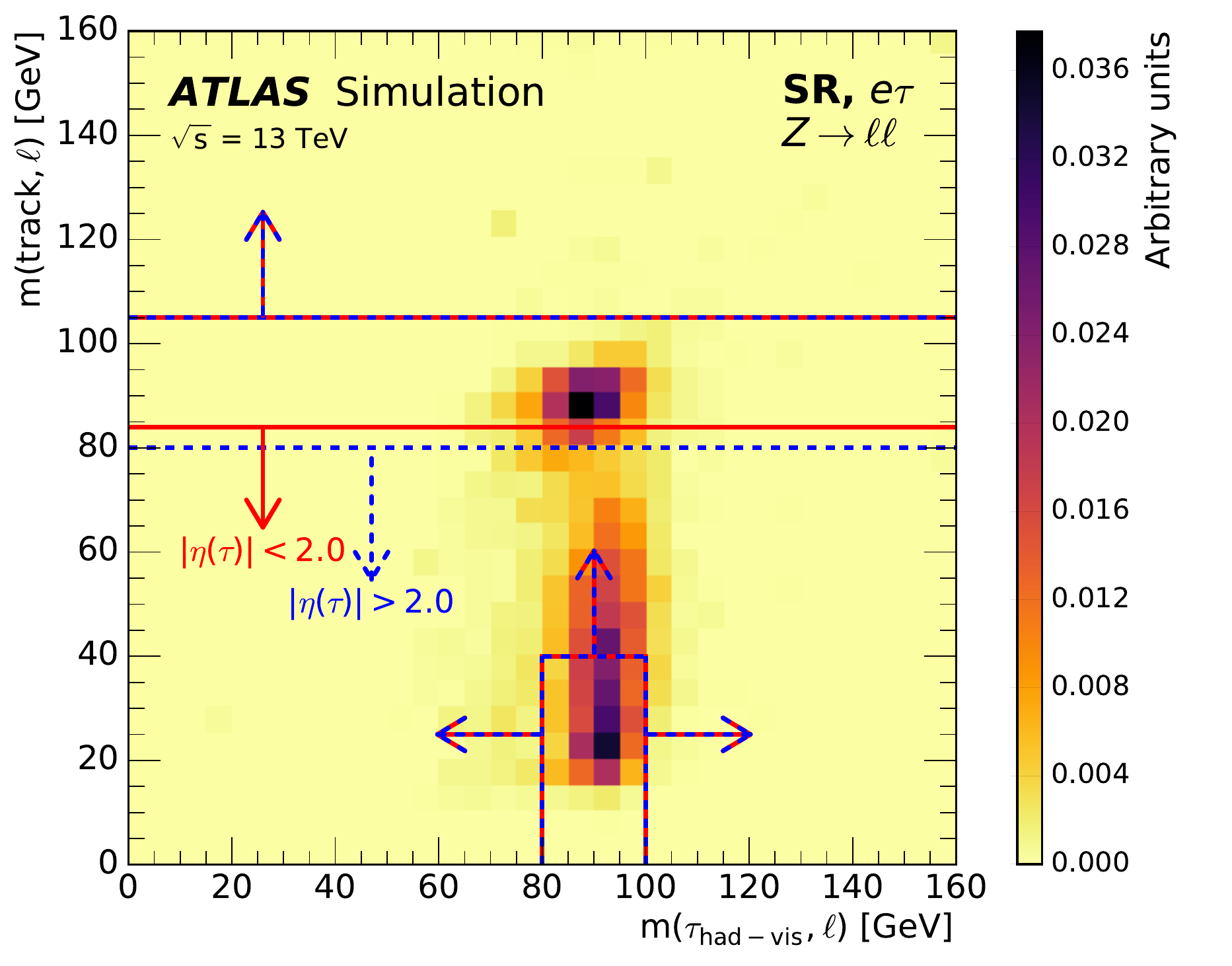}
	{fig:LFVZ_cuts_etau}
	{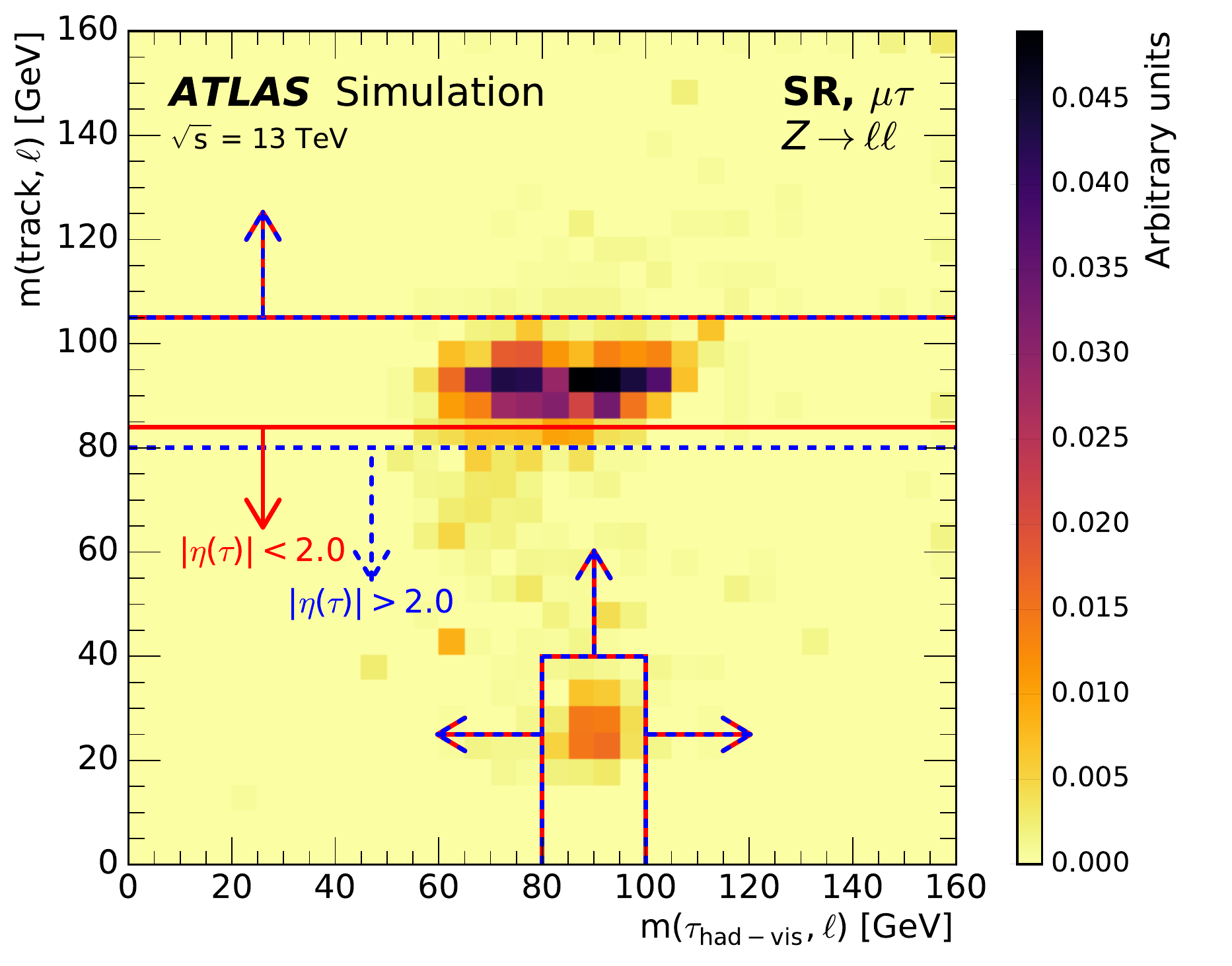}
	{fig:LFVZ_cuts_mutau}
	{Invariant mass of the track-lepton system versus the invariant mass of the $\tau$ candidate-lepton system for the (\subref{fig:LFVZ_cuts_etau}) $e\tau$ and (\subref{fig:LFVZ_cuts_mutau}) $\mu\tau$ channels~\cite{LFVZ}.}
	{fig:LFVZ_cuts}
\doublefig
	{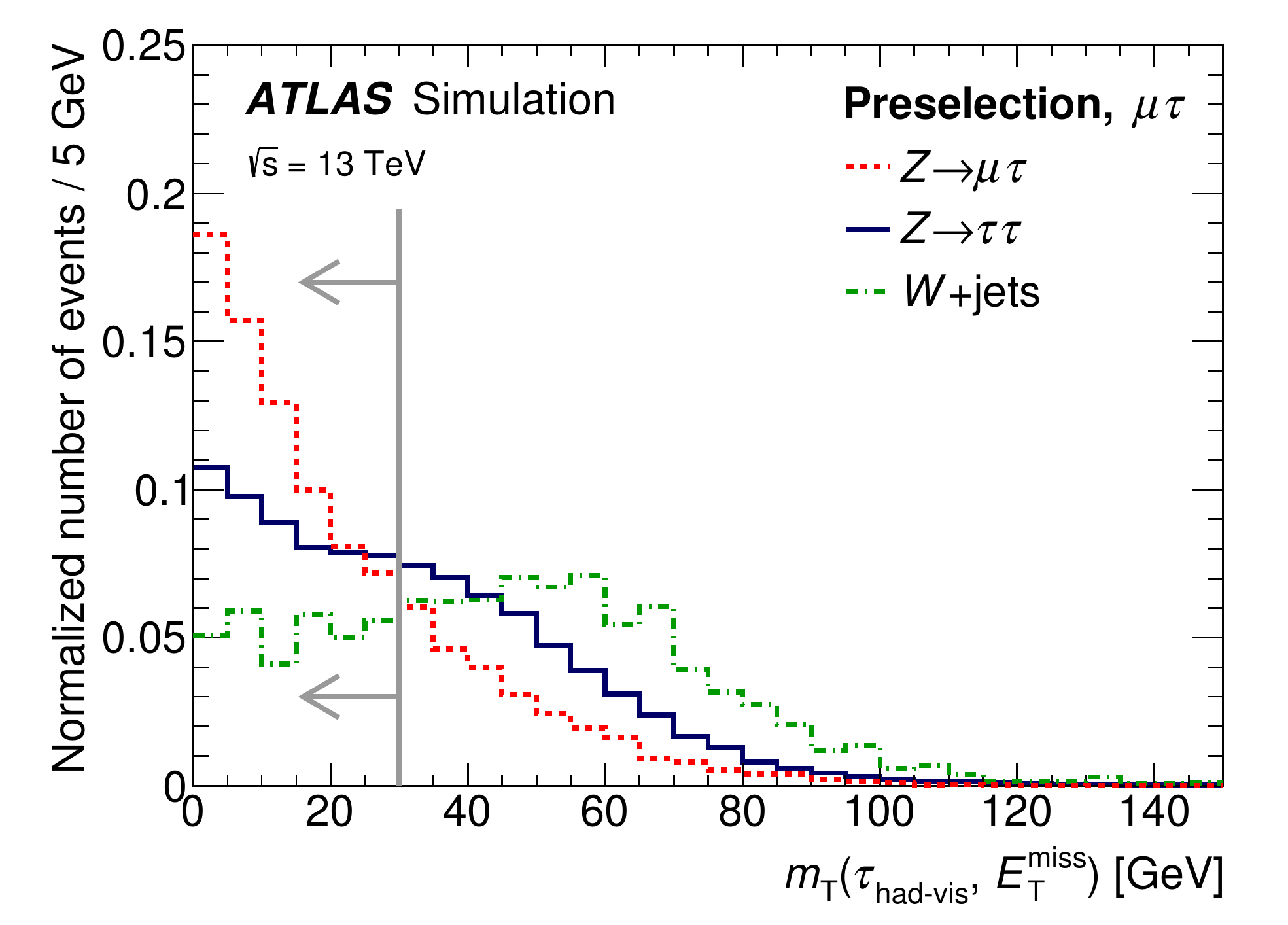}
	{fig:LFVZ_mT_mutau}
	{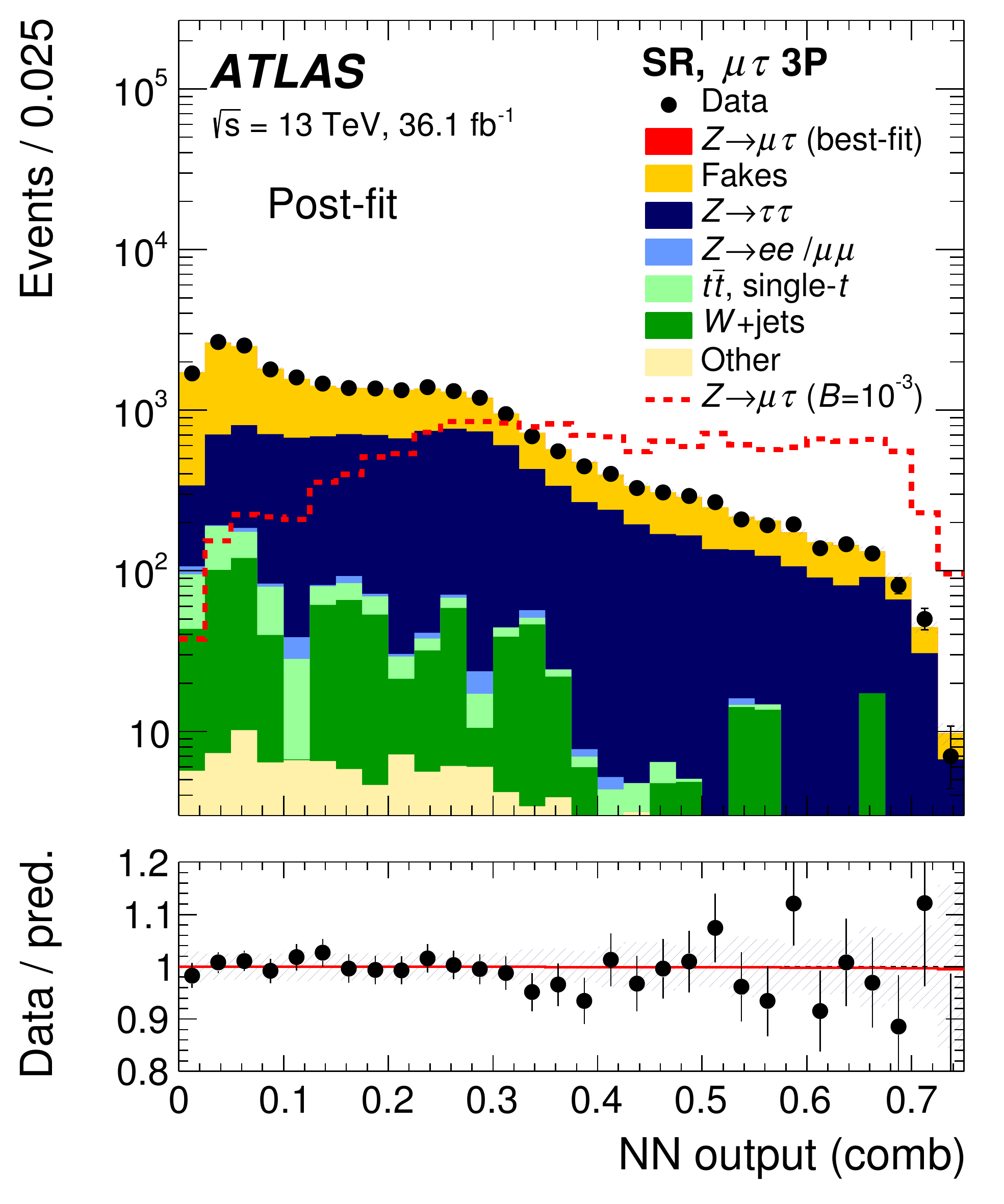}
	{fig:LFVZ_NNout_mutau}
	{(\subref{fig:LFVZ_mT_mutau}) Transverse mass distribution expected for background and signal events. (\subref{fig:LFVZ_NNout_mutau}) NN output for the $\mu\tau$ channel.~\cite{LFVZ}}
	{fig:LFVZ_mT_NNout}

\newpage
\section{Conclusion}
The ATLAS experiment has a rich program of new physics searches using leptonic final states. Four of these searches were presented using the latest recorded data with a center-of-mass energy of 13 \TeV{}, and many were not covered in this text. No significant excess above the SM expectation was found, therefore limits were set on relevant BSM theory parameters. As ATLAS approaches the recording of the full Run 2 dataset, updated results for these and other searches will further increase the sensitivity of ATLAS searches for new physics.

\end{document}